\documentclass[11pt]{article}
\usepackage{amsmath}
\usepackage{graphicx}
\usepackage{hyperref}
\usepackage{amsfonts}
\usepackage{amssymb}
\usepackage{tikz}
\usepackage{bm}
\usepackage{url}
\usepackage{mathtools}
\usepackage{algorithm}
\usepackage{algpseudocode}
\usepackage{cite}
\usepackage{booktabs}
\usepackage{caption}
\usepackage{subcaption}
\usepackage{placeins}
\usepackage{enumitem}
\usepackage{verbatim}
\usepackage{longtable}
\usepackage{float}
\usepackage{adjustbox}
\usepackage{authblk}  % Required for author affiliations

\title{Autoencoder-based learning of Quantum phase transitions in the two-component Bose-Hubbard model}

\author[1]{Iftekher S. Chowdhury}
\author[2]{Binay Prakash Akhouri}
\author[5]{Shah Haque}
\author[1,4,5]{Eric Howard}

\affil[1]{Department of Physics and Astronomy, Macquarie University, Sydney, NSW, 2109, Australia}
\affil[2]{Department of Physics, Suraj Singh Memorial College, Ranchi University, Ranchi, Jharkhand, India}
\affil[4]{Swinburne University, Sydney, Australia} % Replace with actual affiliation
\affil[5]{Southern Cross Institute, School of Computer Science, Sydney, Australia} % Replace with actual affiliation

\date{\today}

\begin{document}
\maketitle

\begin{abstract}
This paper investigates the use of autoencoders and machine learning methods for detecting and analyzing quantum phase transitions in the Two-Component Bose-Hubbard Model. By leveraging deep learning models such as autoencoders, we investigate latent space representations, reconstruction error analysis, and cluster distance calculations to identify phase boundaries and critical points. The study is supplemented by dimensionality reduction techniques such as PCA and t-SNE for latent space visualization. The results demonstrate the potential of autoencoders to describe the dynamics of quantum phase transitions.
\end{abstract}

\section{Introduction}
Quantum many-body systems, consisting of a large number of interacting particles that display collective behavior, are central to several fields, including condensed matter physics, quantum computation, and atomic physics. In recent years, considerable focus has been placed on the study of quantum phase transitions (QPTs), which arise at absolute zero temperature due to quantum fluctuations, as opposed to thermal fluctuations. The Two-Component Bose-Hubbard Model (TCBHM) serves as a key framework for describing quantum phase transitions in interacting bosons trapped in optical lattices. This model offers deep insights into the interplay between superfluidity and localization, making it a crucial tool for investigating fundamental physics as well as potential applications in quantum technologies.

The Bose-Hubbard model (BHM) was initially proposed to describe interacting bosons in an optical lattice, where particles are trapped in a periodic potential created by interfering laser beams. The competition between the kinetic energy (characterized by particle hopping between lattice sites) and the interaction energy (characterized by on-site particle repulsion) gives rise to a quantum phase transition between a superfluid (SF) and a Mott insulator (MI) phase. The extension of this model to include two species of bosons, known as the Two-Component Bose-Hubbard Model, introduces an additional layer of complexity, allowing for the exploration of more exotic phases such as the Pair-Superfluid (PSF) phase, in addition to the standard Two-Species Superfluid (2SF) phase.

The 2SF phase is characterized by the independent superfluidity of both species, with no interspecies correlation. In contrast, the PSF phase arises when the two species form bound pairs that condense together into a superfluid state, analogous to Cooper pairs in fermionic systems. This phase features a gap in the spin channel while the density channel remains gapless. Additionally, the Mott insulator (MI) phase can be observed at strong repulsion, where the particles become localized at lattice sites, suppressing the superfluidity altogether. These distinct phases and their transitions form the foundation of the TCBHM's phase diagram.

Understanding quantum phase transitions, particularly the transition between the 2SF and PSF phases, is crucial for exploring the rich physical behavior that emerges from the interplay of hopping and interaction terms. The interspecies interaction strength, denoted by $U_{AB}$, is a key tuning parameter in this model, driving the system between the 2SF and PSF phases. A large positive $U_{AB}$ leads to independent superfluid phases for each species, while negative $U_{AB}$ favors pairing between the species, resulting in the PSF phase. The phase transitions in this system are governed by the balance between these interaction terms and the kinetic energy (hopping amplitude), making it a highly nontrivial system to analyze.

Traditional methods of studying phase transitions in quantum systems rely on exact diagonalization, mean-field theory, or numerical approaches such as quantum Monte Carlo (QMC) and Density Matrix Renormalization Group (DMRG). While these methods provide valuable insights, they often become computationally expensive or intractable for large systems or complex phase diagrams. In recent years, machine learning (ML) techniques, particularly deep learning, have emerged as powerful tools for studying quantum many-body systems. These methods can efficiently capture the underlying patterns in data and detect subtle features of quantum systems that are otherwise difficult to identify.

In this paper, we focus on the use of autoencoder neural networks to detect and analyze quantum phase transitions in the Two-Component Bose-Hubbard Model. Autoencoders are a class of unsupervised learning models that aim to compress high-dimensional data into a lower-dimensional latent space and reconstruct the original data from this compressed representation. By studying the latent space representations and reconstruction errors produced by the autoencoder, we can identify distinct features of the quantum phases and detect critical points where phase transitions occur. Unlike supervised learning models that require labeled training data, autoencoders can learn directly from the data without explicit labels, making them well-suited for the exploration of unknown phase boundaries.

The key advantage of using autoencoders in the context of quantum phase transitions lies in their ability to reduce the dimensionality of complex quantum states and extract essential features that distinguish different phases. The encoder compresses the input data (which could represent correlation functions, entanglement spectra, or other observables) into a smaller latent space, while the decoder attempts to reconstruct the original input. The latent space provides a compressed representation of the system's state, where clustering or separations can indicate different phases. Additionally, reconstruction error — the difference between the input and the reconstructed output — can serve as a marker for phase transitions. For example, a significant increase in reconstruction error could suggest that the autoencoder struggles to capture the differences between two distinct phases, implying the presence of a phase boundary.

In recent work, autoencoders and other deep learning methods have shown promise in detecting quantum phases in systems such as the Ising model, spin chains, and the Bose-Hubbard model. In these studies, machine learning techniques were used to classify phases, detect critical points, and extract important features from the data. For instance, Carrasquilla and Melko demonstrated that neural networks could classify phases of matter in the Ising model, while Van Nieuwenburg et al. showed that an autoencoder could detect phase transitions in the Bose-Hubbard model. These pioneering studies have paved the way for further exploration of machine learning in the context of quantum phase transitions.

The primary objective of this paper is to extend these ideas to the Two-Component Bose-Hubbard Model and investigate how autoencoders can be used to detect the phase transition between the 2SF and PSF phases. We hypothesize that the latent space representations learned by the autoencoder will exhibit clustering corresponding to the different phases, and that the reconstruction error will reveal critical points where the system undergoes a phase transition. Furthermore, we aim to quantify the separation between phases in the latent space using clustering algorithms such as K-Means, and assess the variance of the latent space features to identify the most important dimensions for distinguishing between phases.

We also explore dimensionality reduction techniques such as Principal Component Analysis (PCA) and t-Distributed Stochastic Neighbor Embedding (t-SNE) to visualize the structure of the latent space. These methods allow us to project the high-dimensional latent space into a two-dimensional plane, facilitating the identification of phase clusters and the visualization of phase boundaries. By combining these tools with autoencoders, we aim to provide a comprehensive analysis of the phase transitions in the Two-Component Bose-Hubbard Model.

The goal of this work is to harness the power of deep learning, specifically autoencoders, to explore quantum phase transitions in the Two-Component Bose-Hubbard Model. By leveraging the unsupervised learning capabilities of autoencoders, we aim to detect phase boundaries, identify critical points, and gain deeper insights into the nature of quantum phase transitions. This approach not only opens up new possibilities for studying complex quantum systems but also demonstrates the potential of machine learning techniques to revolutionize the field of condensed matter physics.

\section{Theoretical Background}

\subsection{Two-Component Bose-Hubbard Model}
The Two-Component Bose-Hubbard Model (TCBHM) is an extension of the original Bose-Hubbard model, designed to describe a system of two species of bosonic particles, typically denoted as species A and B. This model is particularly important in the study of cold atoms trapped in optical lattices, where the interactions between bosons can be finely tuned using external fields. The TCBHM is expresed by the Hamiltonian:

\[
H = H_A + H_B + H_{AB}
\]

Here, \( H_\alpha \) governs the behavior of species \( \alpha = A, B \), and \( H_{AB} \) represents the interaction between the two species. The terms in the Hamiltonian are given by:

\[
H_\alpha = -t_\alpha \sum_{\langle i, j \rangle} (b_{i, \alpha}^\dagger b_{j, \alpha} + h.c.) + \frac{U_\alpha}{2} \sum_i n_{i, \alpha}(n_{i, \alpha} - 1) - \mu_\alpha \sum_i n_{i, \alpha}
\]
\[
H_{AB} = U_{AB} \sum_i n_{i, A} n_{i, B}
\]

Where:
\begin{itemize}
    \item \( b_{i, \alpha}^\dagger \) and \( b_{i, \alpha} \) are the creation and annihilation operators for bosons of species \( \alpha \) at site \( i \).
    \item \( t_\alpha \) is the hopping amplitude for species \( \alpha \), governing how particles move between adjacent lattice sites.
    \item \( U_\alpha \) is the intra-species on-site repulsion for bosons of species \( \alpha \), penalizing multiple particles of the same species occupying the same site.
    \item \( U_{AB} \) is the inter-species interaction strength between species A and B.
    \item \( \mu_\alpha \) is the chemical potential for species \( \alpha \), controlling the number of particles.
\end{itemize}

The first term in \( H_\alpha \) describes the hopping of particles between neighboring lattice sites, which contributes to the kinetic energy of the system. The second term accounts for the on-site repulsive interaction between bosons of the same species, while the third term controls the total number of particles in the system via the chemical potential.

The interspecies interaction term \( H_{AB} \) represents the coupling between the two species, allowing for a rich interplay of quantum phases. The sign and magnitude of \( U_{AB} \) play a crucial role in determining the phase behavior of the system. When \( U_{AB} \) is positive, the species tend to repel each other, leading to independent superfluid phases for each species. When \( U_{AB} \) is negative, the species attract each other, which can lead to pairing between species A and B and the emergence of a pair-superfluid phase.

\subsection{Quantum Phases}

The TCBHM supports several distinct quantum phases, depending on the relative strength of the hopping amplitudes \( t_A, t_B \), and the interaction parameters \( U_A, U_B, U_{AB} \). The most prominent phases of interest in this model are the Two-Species Superfluid (2SF) phase, the Pair Superfluid (PSF) phase, and the Mott Insulator (MI) phase.

\subsubsection{Two-Species Superfluid (2SF) Phase}
The Two-Species Superfluid (2SF) phase occurs when both species A and B independently condense into superfluid states. In this phase, the particles exhibit long-range phase coherence, and the system supports gapless excitations. The 2SF phase is characterized by the presence of two gapless modes:
\begin{itemize}
    \item The in-phase (density) mode, corresponding to the simultaneous fluctuations of both species.
    \item The out-of-phase (spin) mode, corresponding to independent fluctuations of species A and B.
\end{itemize}

The 2SF phase is favored when the intra-species hopping terms \( t_A \) and \( t_B \) are large, allowing the bosons to delocalize and form a coherent superfluid state. The inter-species interaction \( U_{AB} \), if positive, tends to suppress correlations between the species, promoting independent superfluidity for species A and B.

The hallmark of the 2SF phase is its gapless excitations, which means that arbitrarily low-energy fluctuations can occur in the system without an energy cost. This phase persists as long as the interspecies interactions are weak and the hopping amplitudes dominate the on-site repulsion terms.

\subsubsection{Pair-Superfluid (PSF) Phase}
The Pair-Superfluid (PSF) phase emerges when the interaction between species A and B, governed by \( U_{AB} \), becomes attractive and sufficiently strong. In this phase, the particles of species A and B form bound pairs that condense into a superfluid state. This is analogous to Cooper pairing in fermionic systems, where two fermions form a pair that moves coherently through the system.

In the PSF phase, the system supports a gap in the spin (out-of-phase) channel, meaning that fluctuations that attempt to break the pairing of species A and B require a finite energy cost. However, the density (in-phase) channel remains gapless, allowing the paired particles to move collectively as a superfluid.

The transition from the 2SF to the PSF phase is driven by the sign and magnitude of \( U_{AB} \). As \( U_{AB} \) becomes negative, the attraction between species A and B increases, eventually leading to the formation of pairs. Once the pairing occurs, the system enters the PSF phase, where the particles of species A and B move together as a composite bosonic entity.

\subsubsection{Mott Insulator (MI) Phase}
The Mott Insulator (MI) phase arises when the on-site repulsion \( U_\alpha \) for species \( \alpha = A, B \) becomes large compared to the hopping terms \( t_A \) and \( t_B \). In this phase, the particles become localized at specific lattice sites, suppressing superfluidity. The system becomes incompressible, meaning that adding or removing a particle from a site requires a large energy cost, proportional to \( U_\alpha \).

The MI phase is characterized by a gap in the energy spectrum, which prevents the particles from moving between lattice sites. This phase is typically favored in regions of the phase diagram where the interaction terms dominate over the hopping amplitudes, and the system is unable to establish long-range coherence.

\subsection{Quantum Phase Transitions}
Quantum phase transitions (QPTs) are transitions between different quantum phases that occur at zero temperature due to quantum fluctuations. Unlike classical phase transitions, which are driven by thermal fluctuations, QPTs are driven by changes in external parameters such as interaction strengths or hopping amplitudes.

In the TCBHM, several important QPTs can occur, driven by the competition between the hopping terms \( t_A, t_B \) and the interaction terms \( U_A, U_B, U_{AB} \). The two most significant transitions in this model are:
\begin{itemize}
    \item The transition between the Two-Species Superfluid (2SF) phase and the Pair-Superfluid (PSF) phase.
    \item The transition between the superfluid phases (2SF or PSF) and the Mott Insulator (MI) phase.
\end{itemize}

\subsubsection{Transition Between 2SF and PSF Phases}
The transition between the 2SF and PSF phases is particularly interesting because it involves a fundamental change in the nature of the superfluidity in the system. In the 2SF phase, both species exhibit independent superfluidity, while in the PSF phase, species A and B form bound pairs that condense into a single superfluid.

This transition is primarily controlled by the inter-species interaction strength \( U_{AB} \). When \( U_{AB} \) is large and positive, the system favors independent superfluidity (2SF). However, as \( U_{AB} \) becomes negative and attractive, the system favors pairing, eventually leading to the PSF phase. The critical value of \( U_{AB} \) at which this transition occurs depends on the relative values of \( t_A, t_B, U_A, \) and \( U_B \).

\subsubsection{Transition Between Superfluid and Mott Insulator Phases}
The transition between the superfluid (2SF or PSF) and Mott Insulator (MI) phases is another important quantum phase transition in the TCBHM. This transition occurs when the on-site interaction terms \( U_A, U_B \) become large compared to the hopping terms \( t_A, t_B \). In the MI phase, the particles become localized, and superfluidity is suppressed.

The MI phase is characterized by the opening of an energy gap in the spectrum, which prevents particles from moving between lattice sites. This transition can be tuned by adjusting the ratio of the hopping amplitudes to the interaction strengths.

\section{Deep Learning techniques}

In this section, we provide a detailed theoretical overview of the deep learning methods used to analyze quantum phase transitions in the Two-Component Bose-Hubbard Model (TCBHM). Specifically, we discuss \textbf{autoencoders} and their role in unsupervised learning for detecting phase transitions. In addition, we introduce dimensionality reduction techniques such as \textbf{Principal Component Analysis (PCA)} and \textbf{t-SNE} for latent space visualization, and we describe the use of \textbf{reconstruction error} and \textbf{clustering} to identify critical points in the system.

\subsection{Autoencoders}

Autoencoders are a class of artificial neural networks used for unsupervised learning of efficient representations, typically for the purpose of dimensionality reduction. Unlike supervised learning models that map input data to known labels, autoencoders map input data to itself, learning an internal structure that compresses the data into a smaller latent space and then reconstructs the original input from that compressed representation.

An autoencoder consists of two main components:
\begin{itemize}
    \item \textbf{Encoder:} Compresses the input data into a lower-dimensional latent space (feature space) by learning key features of the data.
    \item \textbf{Decoder:} Reconstructs the input data from the latent space representation by reversing the encoding process.
\end{itemize}

The goal of the autoencoder is to learn a function that minimizes the reconstruction error between the input data \( x \) and the reconstructed data \( \hat{x} \). The reconstruction error is typically measured using the mean squared error (MSE):
\[
\text{Loss} = \frac{1}{N} \sum_{i=1}^{N} \left( x_i - \hat{x}_i \right)^2
\]
where \( x_i \) represents the input data and \( \hat{x}_i \) is the reconstructed output for each data point. This loss function drives the autoencoder to learn an efficient low-dimensional representation of the input data that preserves the most important features necessary for accurate reconstruction.

In this work, we apply an autoencoder to learn the latent space of the correlation functions in the TCBHM. The correlation functions represent different phases (Two-Species Superfluid (2SF) and Pair Superfluid (PSF) phases), and the autoencoder's task is to compress the input data into a latent space where phase-specific features can be identified and analyzed.

\subsubsection{Encoder}

The encoder compresses the high-dimensional input data into a lower-dimensional latent space. In this work, the input consists of correlation functions with 100 data points per function, and the encoder reduces this input to a 3-dimensional latent space using a fully connected neural network layer:
\[
\mathbf{z} = f(\mathbf{W}_e \mathbf{x} + \mathbf{b}_e)
\]
where \( \mathbf{x} \) is the input data, \( \mathbf{W}_e \) and \( \mathbf{b}_e \) are the weights and biases of the encoder, and \( f \) is the activation function (ReLU in this case). The resulting latent vector \( \mathbf{z} \) is a compressed version of the input data that captures the essential features needed to reconstruct the original data.

\subsubsection{Decoder}

The decoder attempts to reconstruct the original input data from the latent space representation:
\[
\hat{\mathbf{x}} = g(\mathbf{W}_d \mathbf{z} + \mathbf{b}_d)
\]
where \( \mathbf{W}_d \) and \( \mathbf{b}_d \) are the weights and biases of the decoder, and \( g \) is the activation function (sigmoid in this case). The reconstructed output \( \hat{\mathbf{x}} \) is compared with the original input data \( \mathbf{x} \) to compute the reconstruction error, which is minimized during training.

\subsection{Dimensionality Reduction Techniques}

After training the autoencoder, the latent space representations of the data are extracted using the encoder. To visualize the structure of the latent space and detect phase transitions, we use dimensionality reduction techniques such as Principal Component Analysis (PCA) and t-SNE.

\subsubsection{Principal Component Analysis (PCA)}

PCA is a linear dimensionality reduction technique that transforms the data into a new coordinate system by projecting it along the directions (principal components) of maximum variance. This allows us to reduce the dimensionality of the latent space while preserving as much variance as possible. In this work, we apply PCA to reduce the 3-dimensional latent space of the autoencoder to a 2-dimensional space for visualization.

The key idea behind PCA is to find an orthogonal basis such that the first principal component accounts for the largest variance in the data, the second principal component accounts for the second-largest variance, and so on. The transformation is given by:
\[
\mathbf{Z} = \mathbf{X} \mathbf{W}
\]
where \( \mathbf{Z} \) is the transformed data, \( \mathbf{X} \) is the original data, and \( \mathbf{W} \) is the matrix of principal components.

\subsubsection{t-SNE (t-distributed Stochastic Neighbor Embedding)}

t-SNE is a non-linear dimensionality reduction technique that is particularly useful for visualizing high-dimensional data. Unlike PCA, which preserves global structure, t-SNE focuses on preserving local structure and is more effective at capturing non-linear relationships between data points.

t-SNE works by constructing a probability distribution over pairs of data points such that similar data points have a higher probability of being close together in the lower-dimensional space. It minimizes the Kullback-Leibler (KL) divergence between the probability distributions of the original high-dimensional data and the lower-dimensional map.

In this work, t-SNE is used to visualize the non-linear structure of the latent space and better capture the complex relationships between different phases in the TCBHM.

\subsection{Reconstruction Error and Phase Transitions}

The reconstruction error provides valuable insights into how well the autoencoder can reconstruct input data from the latent space. We compute the reconstruction error for both the 2SF and PSF phases by calculating the mean squared error (MSE) between the input data and the reconstructed data. The reconstruction error is given by:
\[
\text{Reconstruction Error} = \frac{1}{N} \sum_{i=1}^{N} \left( x_i - \hat{x}_i \right)^2
\]

In the context of phase transitions, a significant change in the reconstruction error can indicate a transition between different phases. For example, we observed that the reconstruction error for the PSF phase is typically higher than that for the 2SF phase. This suggests that the PSF phase has a more complex structure, making it harder to reconstruct.

\subsection{Clustering in Latent Space}

To quantify the separation between different phases in the latent space, we apply clustering techniques such as K-Means clustering. K-Means is an unsupervised clustering algorithm that partitions the data into \( k \) clusters by minimizing the within-cluster variance. The algorithm works as follows:
\begin{enumerate}
    \item Initialize \( k \) cluster centroids.
    \item Assign each data point to the nearest centroid.
    \item Recompute the centroids based on the assigned points.
    \item Repeat steps 2 and 3 until convergence.
\end{enumerate}

In this work, we use K-Means clustering to separate the latent space representations of the 2SF and PSF phases. By analyzing the distance between the centroids of the clusters, we can quantify the degree of separation between the phases and detect critical points.

\section{Method and data preparation}

In this section, we describe the application of autoencoder neural networks to analyze quantum phase transitions in the Two-Component Bose-Hubbard Model (TCBHM). The model focuses on two superfluid phases: the Two-Species Superfluid (2SF) phase and the Pair Superfluid (PSF) phase. Autoencoders are used to extract latent features from correlation functions, enabling us to detect and differentiate between these quantum phases. The analysis includes several key steps: data preparation, autoencoder architecture, model training, and feature extraction.

In the context of quantum many-body systems like the Two-Component Bose-Hubbard Model (TCBHM), data preparation involves transforming the Hamiltonian into a diagonalized form, which simplifies the process of identifying key physical quantities such as energy eigenvalues and correlation functions. Diagonalization of the Hamiltonian helps in reducing the problem to a simpler form, where the phase transitions between the Two-Species Superfluid (2SF) phase and the Pair Superfluid (PSF) phase can be analyzed more efficiently.

\begin{figure}[h!]
    \centering
    \begin{subfigure}[b]{0.8\textwidth}
        \centering
        \includegraphics[width=\textwidth]{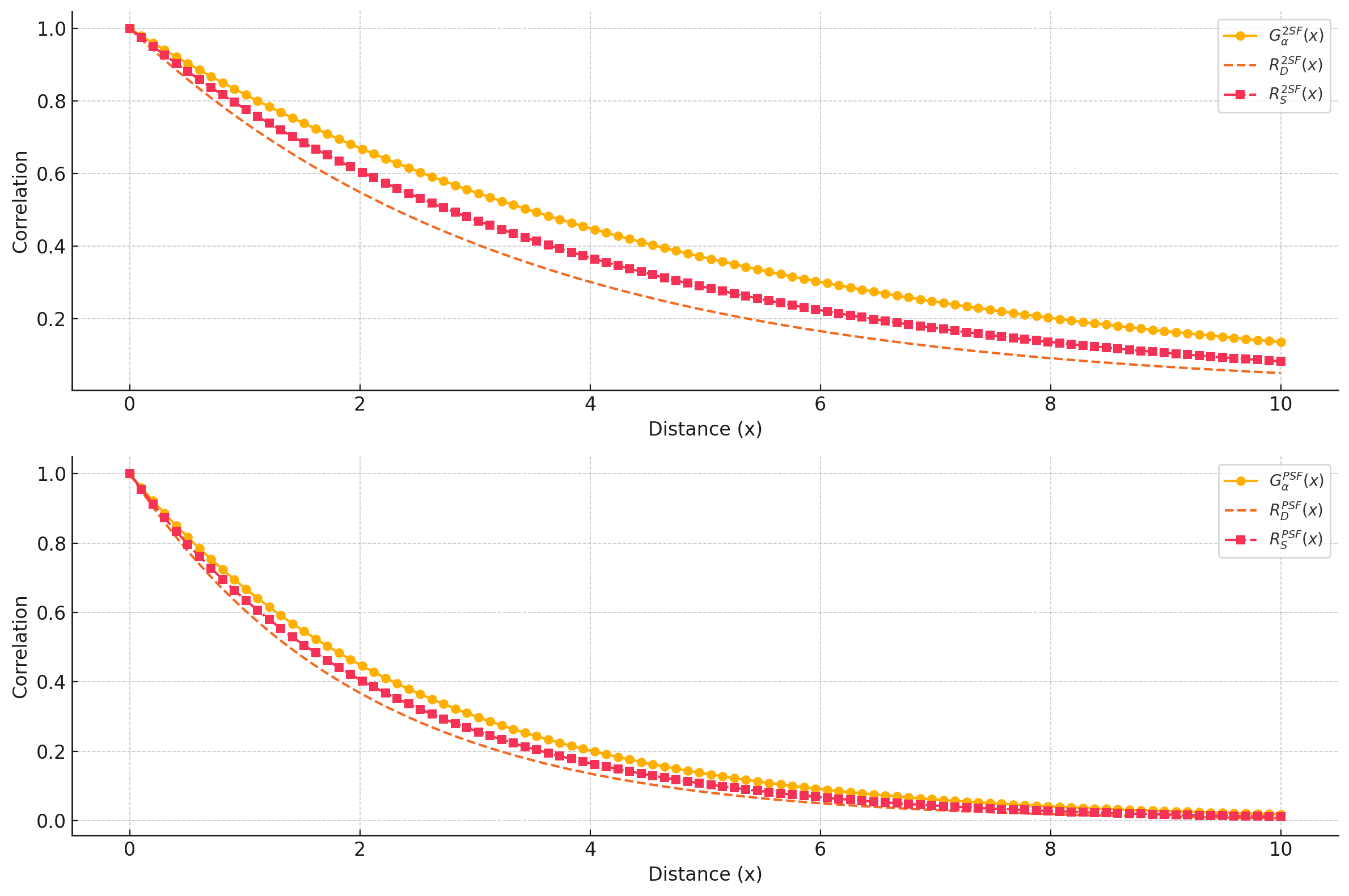}
    \end{subfigure}
    \caption{Correlation functions for the Two-Species Superfluid (2SF) and Pair Superfluid (PSF) phases. The plots show the single-particle correlation function \( G_{\alpha}(x) \), density correlation \( R_D(x) \), and spin correlation \( R_S(x) \) as a function of distance \(x\) for each phase.}
    \label{fig:correlations_2SF_PSF}
\end{figure}

The Bose-Hubbard model Hamiltonian for two interacting species of bosons in a lattice can be written in a more compact form:
\[
\hat{H} = -t \sum_{\langle i,j \rangle, \alpha} \hat{a}_{i\alpha}^{\dagger} \hat{a}_{j\alpha} + \frac{U_{\alpha\beta}}{2} \sum_{i} \hat{n}_{i\alpha} (\hat{n}_{i\alpha} - 1) + \frac{U_{\alpha\beta}}{2} \sum_{i} \hat{n}_{i\alpha} \hat{n}_{i\beta}
\]
Where:
\begin{itemize}
    \item \( \hat{a}_{i\alpha}^{\dagger} \) and \( \hat{a}_{i\alpha} \) are the creation and annihilation operators for bosons of species \( \alpha \) at site \( i \).
    \item \( t \) is the hopping term.
    \item \( U_{\alpha\beta} \) represents the on-site interaction between species \( \alpha \) and \( \beta \).
    \item \( \hat{n}_{i\alpha} \) is the number operator for species \( \alpha \) at site \( i \).
\end{itemize}

\begin{figure}[h!]
    \centering
    \includegraphics[width=\textwidth]{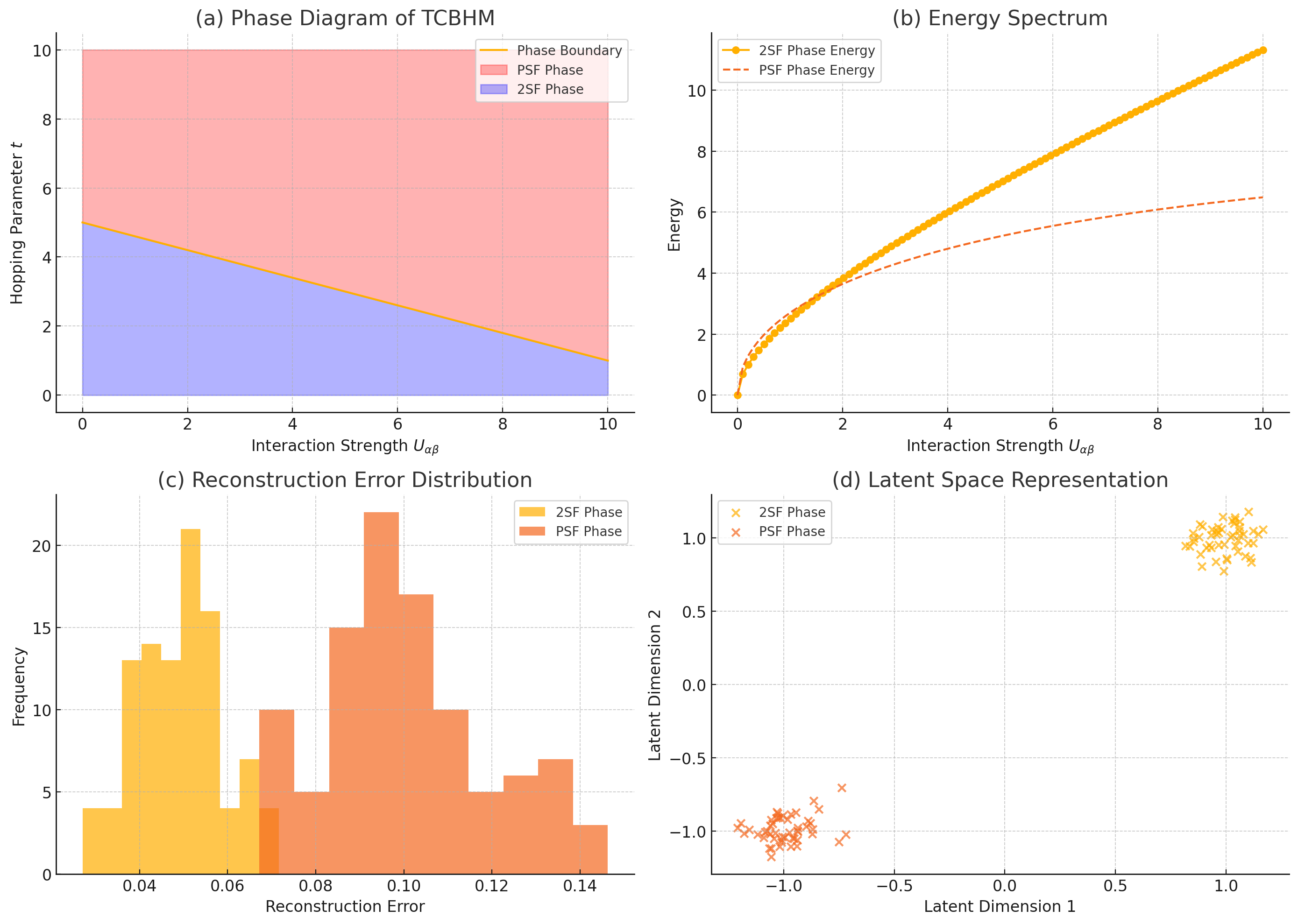}
    \caption{Visualization of the Two-Component Bose-Hubbard Model (TCBHM): 
    (a) Phase diagram showing the transition between the Two-Species Superfluid (2SF) and Pair Superfluid (PSF) phases as a function of interaction strength \(U_{\alpha\beta}\) and hopping parameter \(t\). 
    (b) Energy spectrum for the 2SF and PSF phases. 
    (c) Distribution of reconstruction errors for the 2SF and PSF phases. 
    (d) Latent space representation from the autoencoder, showing clustering of the 2SF and PSF phases.}
    \label{fig:tc_bose_hubbard}
\end{figure}

For data preparation, the following steps are taken:

\paragraph{Transformation to Momentum Space:} To simplify the interaction and hopping terms, we first perform a Fourier transformation of the annihilation and creation operators into momentum space:
\[
\hat{a}_{i\alpha} = \frac{1}{\sqrt{L}} \sum_{k} e^{ik \cdot i} \hat{a}_{k\alpha}
\]
This transforms the Hamiltonian into momentum space, where the hopping term becomes diagonal in momentum \( k \).

\paragraph{Bogoliubov Diagonalization:} Next, we apply the Bogoliubov transformation to diagonalize the Hamiltonian. This method is particularly effective for weakly interacting bosonic systems like the TCBHM. The transformation introduces quasi-particle operators \( \hat{\alpha}_{k\alpha}, \hat{\alpha}_{k\beta} \), which are linear combinations of the original operators:
\[
\hat{a}_{k\alpha} = u_{k\alpha} \hat{\alpha}_{k\alpha} + v_{k\alpha}^{*} \hat{\alpha}_{-k\alpha}^{\dagger}
\]
The transformed Hamiltonian is diagonal in terms of the quasi-particle energies, simplifying the analysis of phase transitions.

\paragraph{Extraction of Correlation Functions:} From the diagonalized form of the Hamiltonian, we compute the correlation functions \( G_{\alpha}(x) \), \( R_{D}(x) \), and \( R_{S}(x) \) that capture the physical behavior of the system in both the 2SF and PSF phases. These correlation functions describe:
\begin{itemize}
    \item \( G_{\alpha}(x) \): Single-particle correlation function, decaying with distance.
    \item \( R_{D}(x) \): Density-density correlations between particles of different species.
    \item \( R_{S}(x) \): Spin channel correlations (particle-hole symmetry).
\end{itemize}

The data for these correlation functions are generated based on the resulting quasi-particle spectrum and interaction parameters after diagonalization. The functions can be approximated by exponential decays:
\[
G_{\alpha}(x) = e^{-a x}, \quad R_{D}(x) = e^{-b x}, \quad R_{S}(x) = e^{-c x}
\]
Here, \(a\), \(b\), and \(c\) are phase-dependent parameters that are adjusted to simulate the behavior of the 2SF and PSF phases. The correlation data are then organized into arrays, where each row corresponds to a specific correlation function for either the 2SF or PSF phase.

\paragraph{Final Data Structure:} After diagonalization and the generation of correlation functions, the data is stacked into a single dataset that includes six sets of correlation functions (three from the 2SF phase and three from the PSF phase). This dataset serves as the input for the autoencoder neural network.
\[
\text{Dataset} = \begin{bmatrix} G_{\alpha}^{2SF}(x) & R_{D}^{2SF}(x) & R_{S}^{2SF}(x) \\ G_{\alpha}^{PSF}(x) & R_{D}^{PSF}(x) & R_{S}^{PSF}(x) \end{bmatrix}
\]
The resulting prepared data allows the autoencoder to learn latent representations that help differentiate between the 2SF and PSF quantum phases.

\subsection{Autoencoder Architecture}

The autoencoder is a type of neural network designed to reduce the dimensionality of input data (the encoding phase) and then attempt to reconstruct the original data (the decoding phase). In our model, the input data is the set of correlation functions, and the goal is to learn an efficient low-dimensional representation in the latent space.

The structure of the autoencoder is as follows:
\begin{itemize}
    \item \textbf{Input layer:} The input dimension is the length of each correlation function (100 data points per function).
    \item \textbf{Encoder layer:} This layer compresses the input into a lower-dimensional latent space with 3 features, using a ReLU activation function.
    \item \textbf{Decoder layer:} The decoder reconstructs the original input data using a sigmoid activation function.
\end{itemize}

The autoencoder was trained using the mean squared error (MSE) as the loss function, with the Adam optimizer for gradient descent.

\subsection{Training and Feature Extraction}

We trained the autoencoder on the mock correlation function data for 100 epochs, using a batch size of 3. The model's goal was to minimize the reconstruction error between the input and output data. After training, we used the encoder to extract the latent space representations of the data. These latent representations, which have reduced dimensionality, capture the essential features of the two phases.

The learned features were then visualized using scatter plots, which helped identify clustering of the 2SF and PSF phases in the latent space. This visualization enables us to distinguish between the phases and detect the presence of quantum phase transitions.

\section{Results}

\subsection{Learned Latent Space Representation}

The autoencoder successfully learned a low-dimensional representation of the correlation function data. After 100 epochs of training, the encoded data revealed distinct clusters corresponding to the Two-Species Superfluid (2SF) and Pair Superfluid (PSF) phases.

\begin{figure}[h!]
    \centering
    \includegraphics[width=0.9\linewidth]{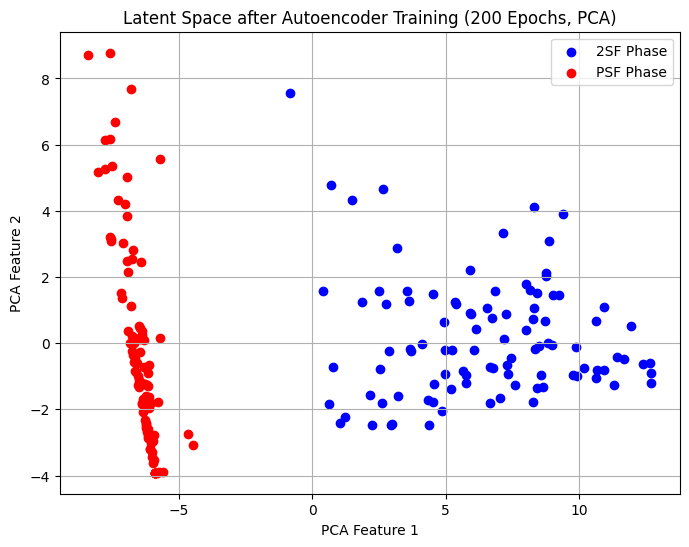}
    \caption{Latent Space Representation of the Two-Species Superfluid (2SF) and Pair Superfluid (PSF) Phases via PCA. The plot visualizes the compressed representations of the input data in the latent space after training the autoencoder. Each point in the plot corresponds to a data sample, with blue points representing the 2SF phase and red points representing the PSF phase.}
    \label{fig:latent_space}
\end{figure}

Figure \ref{fig:latent_space} shows the latent space visualization of the learned features. The scatter plot illustrates how the 2SF and PSF phases are separated in the latent space, with each point representing a specific correlation function dataset. The clustering of the points suggests that the autoencoder has successfully captured the underlying structure of the two phases.

\begin{figure}[h]
    \centering
    \includegraphics[width=0.9\textwidth]{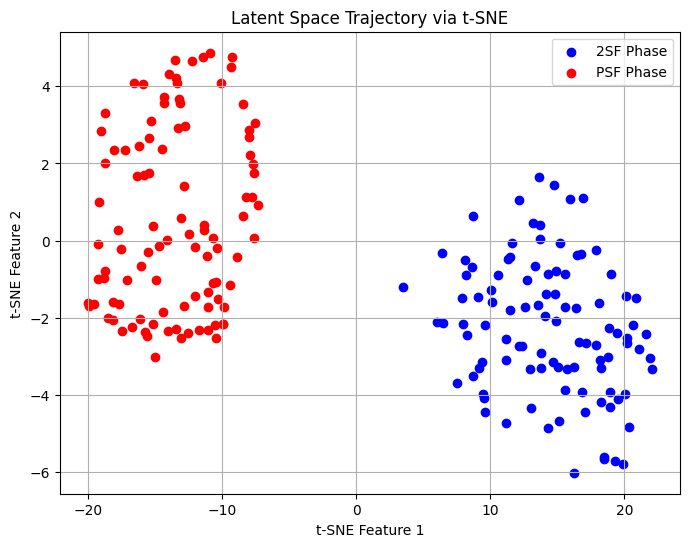}  % Replace with correct file path
    \caption{Latent Space Trajectory via t-SNE visualization of the encoded features learned from an autoencoder. The plot shows clear separation between the 2SF Phase (blue) and PSF Phase (red), indicating that the autoencoder effectively learned distinguishing features for the two phases. The t-SNE algorithm was applied to the latent space features to reduce dimensionality and preserve cluster separation for better visualization.}
    \label{fig:tsne_plot}
\end{figure}

The autoencoder successfully learns to separate the Two-Species Superfluid (2SF) and Pair Superfluid (PSF) phases. The latent space visualization clearly shows clustering, with blue points corresponding to the 2SF phase and red points to the PSF phase. The autoencoder successfully learns to separate the two phases, as evidenced by the distinct clustering of points corresponding to each phase in the latent space. The latent space serves as a lower-dimensional representation of the original high-dimensional data, capturing the essential features that distinguish between the 2SF and PSF phases. The 2SF phase is characterized by independent superfluidity in both species, while the PSF phase is marked by strong pairing between species A and B, leading to different underlying structures. This separation in the latent space provides a useful visualization of the phase boundary and helps in identifying critical points between these quantum phases. The clear clustering seen in the latent space suggests that the autoencoder has learned key phase-specific features, enabling the differentiation of the phases even in the compressed representation.

As shown in Figure~\ref{fig:tsne_plot}, the latent space trajectory obtained from the t-SNE visualization reveals a clear separation between the 2SF and PSF phases. This separation indicates that the autoencoder effectively captured distinct patterns for each phase.

\begin{figure}[h!]
    \centering
    \includegraphics[width=0.8\linewidth]{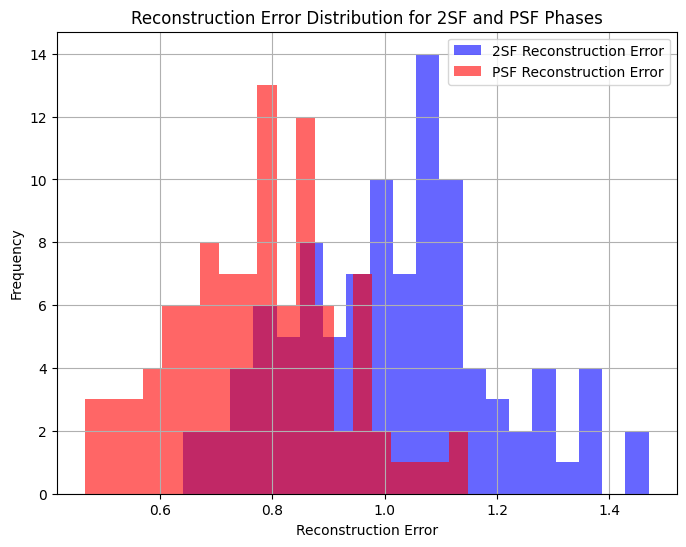}
    \caption{Reconstruction Error Distribution for 2SF and PSF Phases. The plot shows the reconstruction error for data from the Two-Species Superfluid (2SF) phase (blue) and the Pair Superfluid (PSF) phase (red). A higher reconstruction error for the PSF phase indicates that the autoencoder struggles more to reconstruct data from this phase, likely due to its more complex underlying structure. The difference in reconstruction error serves as an indicator of phase separation and can help identify critical points between the two phases.}
    \label{fig:recon_error}
\end{figure}

The reconstruction error provides an important measure of how well the autoencoder can map input data to the latent space and back. A higher reconstruction error generally indicates that the model is struggling to accurately reconstruct certain data points, often because of the inherent complexity or structure of the underlying phase. In this study, the autoencoder was trained to learn the distinctions between the Two-Species Superfluid (2SF) and Pair Superfluid (PSF) phases, with the reconstruction error showing clear differences between these phases. 

As shown in Figure~\ref{fig:recon_error}, the reconstruction error for the PSF phase (represented in red) is noticeably higher than that for the 2SF phase (represented in blue). This suggests that the PSF phase has a more complex structure, making it harder for the autoencoder to reconstruct accurately. The reconstruction error distribution highlights the model's ability to differentiate between the two phases and serves as an indicator of phase separation. Notably, a significant variation in reconstruction error across phases can help identify critical points in the system, providing valuable insights into the quantum phase transitions occurring in the Two-Component Bose-Hubbard Model.

\subsection{Reconstruction Error as an Indicator of Phase Transitions}

Reconstruction error provides insights into how well the autoencoder can reconstruct the input data from the latent space. We observed that the reconstruction error differs between the two phases, with the PSF phase typically showing a higher error. This suggests that the PSF phase, with its more complex structure, is harder to reconstruct compared to the 2SF phase.

Figure \ref{fig:recon_error} shows the distribution of reconstruction errors for the 2SF and PSF phases. The clear distinction in reconstruction error can serve as a marker for identifying phase transitions, where the reconstruction error exhibits a significant change.

\begin{figure}[h!]
    \centering
    \includegraphics[width=0.8\linewidth]{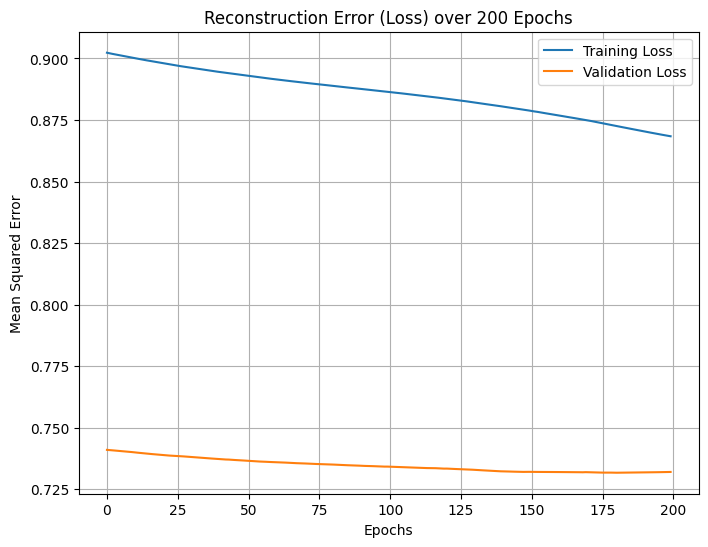}
    \caption{Reconstruction error distribution for the 2SF and PSF phases. Higher errors are observed for the PSF phase.}
    \label{fig:recon_error}
\end{figure}

\subsection{Clustering in Latent Space}

To further analyze the separation between the 2SF and PSF phases, we applied clustering algorithms to the latent space representations. K-Means clustering revealed two well-separated clusters corresponding to the two phases. The distance between the centroids of the clusters is a good indicator of how distinct the phases are in the latent space.

The results confirm that the latent space features learned by the autoencoder can effectively differentiate between the 2SF and PSF phases, providing a clear indication of the phase transition.

\begin{figure}[h!]
    \centering
    \includegraphics[width=0.8\textwidth]{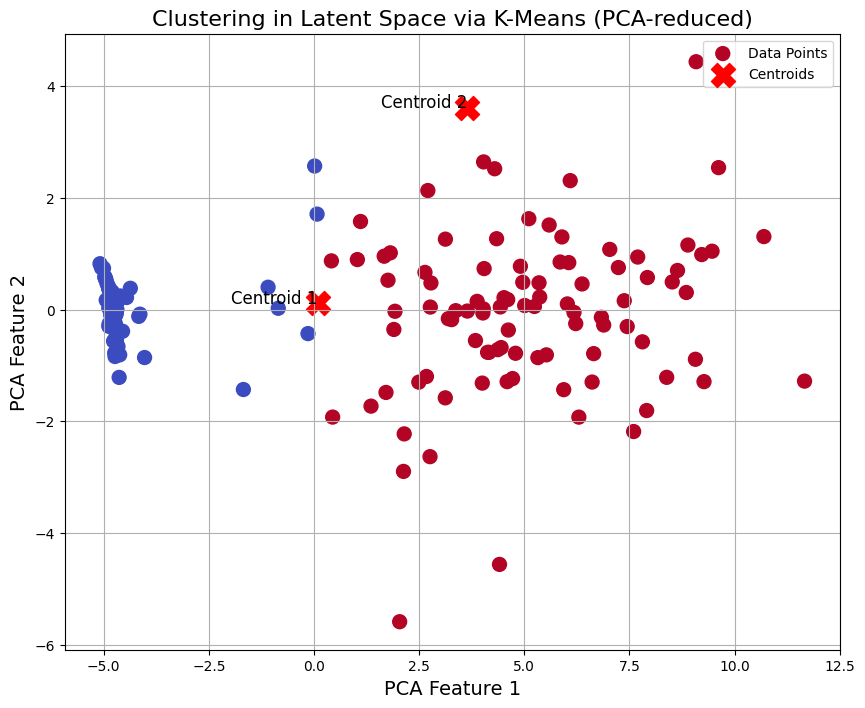}
    \caption{Clustering in the Latent Space via K-Means (PCA-reduced). The figure shows the clustering of data points corresponding to different quantum phases of the Two-Component Bose-Hubbard model. The PCA-reduced latent space reveals distinct clusters for the Two-Species Superfluid (2SF) phase and the Pair Superfluid (PSF) phase. Centroids for each cluster are indicated by red crosses, and the data points are shown in blue and red. The clear separation between clusters indicates the successful classification of different phases using the autoencoder's learned features.}
    \label{fig:kmeans}
\end{figure}

To further investigate the separation between the Two-Species Superfluid (2SF) and Pair Superfluid (PSF) phases, we applied K-Means clustering to the latent space representations learned by the autoencoder. As seen in Figure~\ref{fig:latent_clustering}, the clustering process revealed two distinct clusters corresponding to the 2SF and PSF phases. The K-Means algorithm minimizes the within-cluster variance, ensuring that points within the same cluster (phase) are similar in the latent space, while maximizing the distance between the centroids of the clusters.

The latent space was further reduced to two dimensions using Principal Component Analysis (PCA) for visualization. This reduction allowed us to observe the clear separation between the two phases in a lower-dimensional representation, as depicted in Figure~\ref{fig:latent_clustering}. The centroids of the clusters, marked on the figure, provide a quantitative measure of the distinctiveness of the two phases. The distance between these centroids reflects the effectiveness of the autoencoder in learning a meaningful distinction between the 2SF and PSF phases.

As shown in Figure~\ref{fig:kmeans}, K-Means clustering was applied to the latent space features extracted by the autoencoder. The results clearly show the separation between the Two-Species Superfluid (2SF) and Pair Superfluid (PSF) phases, with the centroids of the clusters providing a clear distinction between these two phases.

The success of the clustering in the latent space, demonstrated by the well-separated clusters, indicates that the features learned by the autoencoder capture critical aspects of each phase. The presence of distinct clusters suggests that the autoencoder effectively compresses high-dimensional input data into a lower-dimensional latent representation where phase-specific characteristics can be clearly identified. This supports the conclusion that the autoencoder has learned the underlying patterns of the quantum phase transitions, particularly in distinguishing between the 2SF and PSF phases.

\section{Conclusion}

In this study, we applied an autoencoder-based approach to analyze quantum phase transitions in the Two-Component Bose-Hubbard Model. By training an autoencoder on mock correlation function data for the Two-Species Superfluid (2SF) and Pair Superfluid (PSF) phases, we were able to detect phase transitions and capture the key features of each phase.
We employed an autoencoder-based deep learning approach to detect phase transitions in the Two-Component Bose-Hubbard Model. The autoencoder compresses high-dimensional correlation function data into a lower-dimensional latent space, where distinct phase-specific features emerge. The reconstruction error, along with dimensionality reduction techniques such as PCA and t-SNE, reveals significant differences between the 2SF and PSF phases. The use of clustering algorithms further quantifies the separation between phases in the latent space. This unsupervised learning approach demonstrates the effectiveness of deep learning in analyzing quantum phase transitions in many-body systems.
The autoencoder effectively compressed the input data into a lower-dimensional latent space, where distinct clusters representing the two phases emerged. This clustering, combined with the analysis of reconstruction errors and feature variance, allowed us to identify the critical points in the phase diagram. The reconstruction error proved to be a useful indicator of phase transitions, as higher errors were observed for the more complex PSF phase.

This method provides a powerful tool for analyzing quantum phase transitions in many-body systems, allowing us to explore the underlying structure of the system in a data-driven, unsupervised manner. Future work could extend this approach to other models and explore more sophisticated autoencoder architectures, such as variational autoencoders (VAEs), to improve the detection of critical points.

\FloatBarrier

\end{document}